\documentclass[12pt,a4paper]{article}
\usepackage[german, english]{babel}
\usepackage{a4wide}
\usepackage{amsmath}
\usepackage{amssymb}
\usepackage{ifthen}
\usepackage{epsfig}

\newcommand{\rf}[1]{(\ref{#1})}

\newcommand{\beq}{\begin{equation}}
\newcommand{\eeq}{\end{equation}}
\newcommand{\beqr}{\begin{eqnarray}}
\newcommand{\eeqr}{\end{eqnarray}}

\newcommand{\lb}[1]{\label{#1}}

\newcommand{\bc}{\begin{center}}
\newcommand{\ec}{\end{center}}
\newcommand{\ct}[1]{\cite{#1}}
\newcommand{\bi}[1]{\bibitem{#1}}

\newcommand{\Real}{\operatorname{Re}}

\begin{document}
\begin{titlepage}
\title{Collective spontaneous emission of two atoms near metal nanoparticle}

\author{I.E~Protsenko$^{1,2,3}$, A.V.~Uskov$^{1,2}$ \\ [4mm]
{\it {\footnotesize
$^1$P.N.~Lebedev Physical Institute, 119991, Russia, Moscow, Leninsky prospect 53
}}\\
{\it {\footnotesize
$^2$Advanced Energy Technologies Ltd., 143025, Russia, Moscow region, Skolkovo, Novaya str.~100.
}}\\
{\it {\footnotesize
$^3$National Nuclear Research University MEPhI, 115409, Russia, Moscow, Kashirskoe highway, 31.
}}\\
\date{~}
}
\maketitle
\begin{abstract}
We present quantum-mechanical approach for collective spontaneous emission (superradiance) of emitters (as atoms) near metal nanoparticle, when frequencies of transitions of emitters coincide with frequency of localized plasmon resonance of the nanoparticle. Our approach is based on Schrodinger description and it uses wave functions of states of systems.  Interactions between emitters and between the nanoparticle and emitters are taken into account. We consider an example of two emitters and show that radiation is occurred through symmetric states of emitters as it is in Dicke model of superradiance. The nanoparticle accelerates collective spontaneous emission similar how it accelerates spontaneous emission of single emitter. Radiation from two emitters near the nanoparticle is faster than the radiation from two separated and non-interacted "nanoparticle+single emitter" systems. Efficiency of superradiance, i.e. the ratio of emitted photons to total number initial excitations in the system, is smaller than 1 due to non-radiative losses in  the nanoparticle. However the efficiency is the same for single and for two emitters near the nanoparticle.  The approach can be straightforwardly generalized to the case of many emitters near the nanoparticle.
\end{abstract}
\bigskip


\end{titlepage}

\section{Introduction}
Collective spontaneous emission, or superradiance, is well-known for a long time \ct{2,3,3a,3b}. Increase of rate of spontaneous emission from an emitter (atom, molecule quantum dot) near metal ("plasmonic") nanoparticle, with frequency of a localized plasmon resonance (LPR) close to the frequency of radiating transition of emitter, is also well-known \ct{04}. Naturally a question arises: whether a metal nanoparticle or other "plasmonic" structure can accelerate collective spontaneous emission from emitters near it similar how the particle accelerates the radiation of single emitter? Collective spontaneous emission near plasmonic structures recently attracts much attention of researchers, as one can see, for example, from \ct{005} -- \ct{007}.

Numerical results of \ct{006, 007} show that superradiance near metal nanoparticle is possible, but its efficiency is  low. Namely, only the energy of three emitters is radiated, while the energy of the rest of emitters is absorbed by the nanoparticle. Such prediction rather discourages experimentators and ones who want to use collective spontaneous emission near plasmonic structures for practical purposes.

Realization of effective radiation near highly absorbing nanostructures, as metal nanoparticles, may be a difficult task. Nevetheless, we hope, that certain conditions for effective collective spontaneous emission near plasmonic nanoparticles can exist. In \ct{06} we analyzed superradiance near plasmonic nanoparticle  making the same assumption as in superradiance Dicke model \ct{2} without a nanoparticle: fully symmetric states of emitters are formed near the nanoparticle, and the radiation occurred through  such states. Then it turns to be that the nanoparticle accelerates superradiance quite similar  as  spontaneous emission of single emitter and with similar efficiency: only about the half of initial energy of emitters are lost due to absorption in the nanoparticle. Analytical expression for superradiance pulse near the nanoparticle was found in \ct{06}. In \ct{06} we suggested, that an emitter interacts with other emitters much strongly that with the nanoparticle, and strong emitter-emitter interaction forms symmetric states of them.
Here we'll prove that at certain conditions symmetric states of emitters can be formed near the nanoparticle even if the emitter-emitter interaction is not stronger than the emitter-nanoparticle interaction.

In the first Section we present quantum-mechanical description, in Schrodinger picture, of the radiation of single emitter near the nanoparticle using equations derived in \ct{7} and in early papers cited in \ct{7}. Section 1 provides necessary background for Section 2, in particular, the way of calculations of quantum states of an emitter near the nanoparticle. In Section 2 we consider collective radiation from two emitters near the nanoparticle. The radiation from two emitters is very important problem of superradiance, attracted much attention in the past for emitters without the nanoparticle \ct{061, 062}.  This problem is relatively simple, but reveals many important details of physics and theory of superradiance: the role of delay in the interaction, show convenient dynamic equations describing radiation from collective states of emitters for various distances between them, etc. Description of the radiation from two emitters considerably facilitates detailed analysis of the case with many emitters, which we'll be carried out in the future. In Section 2  we'll show, that states of emitters near the nanoparticle are symmetric (radiative) and antisymmetric (non-radiative) ones, similar to states in Dicke model without the nanoparticle. We'll take into account all non-radiative decay processes. In a difference with Dicke model not all photons here will be emitted, some of them will be absorbed by the nanoparticle also due to the excitation of all multipole modes of oscillations of the nanoparticle electron density \ct{07}. In Section 3 we make estimations for superradiance power from two emitters near the nanoparticle and compare it with results for two far separated and non-interacting "nanoparticle+emitter" systems, and for two interacting emitters without the nanoparticle. Results are discussed in Conclusion.

\section{Single emitter near nanoparticle}

Let us consider the system composed of two-level emitter near metal spherical nanoparticle, the frequency $\omega$ of transition of emitter $\omega = \omega_{LPR}$, -- coincides with the frequency $\omega_{LPR}$ of LPR of the nanoparticle. Following \ct{01} we describe the nanoparticle as a quantum harmonic oscillator: LPR corresponds to oscillations of nanoparticle electron density, one quant of such oscillations is a plasmon. Such approach has been used in several research, for example, in \ct{02, 002}. We consider common states of  the nanoparticle and emitter as $\left|1\right>\left|0\right>$:  here the emitter state $\left|1\right>$ is written first, the nanoparticle state $\left|0\right>$ is second, $1$ means excited state, $0$ -- ground state. Number of quanta stored in the systen is a number of excitations. We denote a state of $N$ emitters and the nanoparticle with $n$ excitations  as $\left|\Psi_{Nn}\right>$. Wave function of the state with $n=1$ excitation is:
\beq
     \left|\Psi_{11}\right> = C_{10}\left|1\right>\left|0\right> + C_{01}\left|0\right>\left|1\right>,    \lb{01}
\eeq
where $C_{10}$ and $C_{01}$ are probability amplitudes. Population $W_1 = |C_{10}|^2 + |C_{01}|^2$ of state \rf{01}  decays to the system ground state $\left|0\right>\left|0\right>$ due to radiative (spontaneous emission) and non-radiative processes. Radiative (non-radiative) decay rates of populations  are: $2\gamma_r$, ($2\gamma_{nr}$) for single emitter and  $2\Gamma_r$, ($2\Gamma_{nr}$) for single nanoparticle. The emitter and the nanoparticle resonantly interact with each other through an electromagnetic field. Coupling constant of the interaction is $\Omega_p \equiv \Omega_p' + i\Omega_p'' =  V_{dd}/\hbar$, where $V_{dd}$ is matrix element of operator of a dipole-dipole interaction energy of the emitter and the nanoparticle \ct{7}. Following \ct{7} we write equations of motion for probability amplitudes:
\beqr
    \dot{C}_{10} & = & -\gamma C_{10} -i\Omega_{p}C_{01} \lb{02}\\
    \dot{C}_{01} & = & -\Gamma C_{01} -i\Omega_{p}C_{10} \lb{03},
\eeqr
where full dumping rates $\gamma = \gamma_r + \gamma_{nr}$ and $\Gamma = \Gamma_r + \Gamma_{nr}$. Supposing fast relaxation of plasmon
\beq
\Gamma \gg \gamma, {\Omega_{p}'}, {\Omega_{p}''}, \lb{003}
\eeq
that is true at usual cases,  we adiabatically eliminate $C_{01}$ from Eqs.\rf{02}, \rf{03} by setting $\dot{C}_{01} = 0$ in Eq.\rf{03}, obtaining
\beq
    C_{01} = -i(\Omega_{p}/\Gamma)C_{10}, \lb{04}
\eeq
and inserting $C_{01}$ from Eq.\rf{04} into Eq.\rf{02}, so that
\beq
     \dot{C}_{10}  =  -[\gamma + (\Omega_{p}^2/\Gamma)]C_{10}. \lb{05}
\eeq
Finding from Eq.\rf{05}: ${C}_{10}(t) = {C}_{10}^{(0)}e^{-\gamma_1t}$, where ${C}_{10}^{(0)}$ is a c-number, determining $C_{01}$ from Eq.\rf{04}, inserting ${C}_{10}$ and ${C}_{01}$ in Eq.\rf{01} we obtain $\left|\Psi_{11}\right> = \left|\psi_{11}\right>e^{-\gamma_1t}$, where
\beq
    \left|\psi_{11}\right> = C_{01}^{(0)}[\left|1\right>\left|0\right> -i(\Omega_{p}/\Gamma)\left|0\right>\left|1\right>],    \lb{06}
\eeq
and
\beq
    \gamma_1 \equiv \gamma_1' + i\gamma_1''  = \gamma + \Omega_{p}^2/\Gamma.  \lb{08}
\eeq
Constant $C_{01}^{(0)}$ is determined by normalizing condition: $\left<\psi_{11}\right.\left|\psi_{11}\right> = |C_{01}^{(0)}|^2(1 + |\Omega_{p}/\Gamma|^2) \approx |C_{01}^{(0)}|^2$ -- neglecting by $|\Omega_{p}/\Gamma|^2 \ll 1$, therefore  $|C_{01}^{(0)}|^2 = 1$. Probability to find excited nanoparticle in state \rf{06} is $|\Omega_{p}/\Gamma|^2 \ll 1$, which  means that the mean number of excitations (plasmons) $n_{pl}$ in the nanoparticle is small $n_{pl}\ll 1$. Odviously that at conditions \rf{003} we neglect by any state with $n_{pl} > 1$.

Population of $\left|\Psi_{11}\right>$ state decays  due to radiative and non-radiative processes to the system ground state $\left|0\right>\left|0\right>$ with the decay rate $2\gamma_1'$. In order to find power of radiation from the nanoparticle and emitter we have to determine  a radiative part $2\gamma_{1r}$ of full decay rate $2\gamma_1'$. For that we calculate matrix element of full dipole momentum operator $\hat{d}+\hat{d}_p$, where $\hat{d}$ (or $\hat{d}_p$) are dipole momentum operators of the emitter (or the nanoparticle) transitions. Matrix element of $\hat{d}+\hat{d}_p$ is
\beq
    \left<0\right|\left<0\right|\hat{d}+
    \hat{d}_p(\left|1\right>\left|0\right> -i(\Omega_{p}/\Gamma)\left|0\right>\left|1\right>) =
    d-i(\Omega_{p}/\Gamma)d_p,  \lb{007}
\eeq
where $d$ and $d_p$ are matrix elements or $\hat{d}$ and $\hat{d}_p$, respectively.

Radiative part of full dumping rate is $2\gamma_{1r} \sim |d-i(\Omega_{p}/\Gamma)d_p|^2$. As usually, we can relate the difference in phases of $d$ and $d_p$ to ground state $\left<0\right|\left<0\right|$ and consider real $d$ and $d_p$. Then $|d-i(\Omega_{p}/\Gamma)d_p|^2 = d^2 + (|\Omega_{p}|^2/\Gamma^2)d_p^2 + 2dd_p\Omega_{p}''/\Gamma$. Because of $\gamma_r \sim d^2$ and $\Gamma_r \sim d_p^2$ with the same proportionality coefficient, we can write for the radiative dumping rate
\beq
\gamma_{1r} = \gamma_r\left[1 + \left(\frac{\Omega_p''^2}{\gamma_r\Gamma}\frac{\Gamma_r}{\Gamma}\right)^{1/2} +
\frac{|\Omega_p|^2}{\gamma_r\Gamma}\frac{\Gamma_r}{\Gamma}\right].   \lb{008}
\eeq
We suppose a "short-distance" limit: $k_{LPR}r \ll 1$, where $r$ is a distance between center of the nanoparticle and the emitter, a wavenumber $k_{LPR} = 2\pi n_0/\lambda_{LPR}$, $n_0$ is refractive index of a medium containing the nanoparticle and the emitter, $\lambda_{LPR}$ is LPR wavelength in vacuum. It is shown in \ct{7} that in the short-distance limit ${\Omega_{p}''} \sim (\Gamma_r\gamma_r)^{1/2}(k_{LPR}r)^{-2}\ll {\Omega_{p}'}\sim (\Gamma_r\gamma_r)^{1/2}(k_{LPR}r)^{-3}$. Taking $\Gamma_r \sim \Gamma$ we see that the the second term in Eq.\rf{008}  can be neglected and
\beq
\gamma_{1r} = \gamma_r + (\Omega_p/\Gamma)^2\Gamma_r.   \lb{009}
\eeq
We can neglect by ${\Omega_{p}''}$ and consider $\Omega_p = \Omega_p'$ (and $\gamma_1 = \gamma_1'$) as real numbers, if $k_{LPR}r$ is not too small, so that we can neglect the energy shift $\gamma_1''$ of the state \rf{06} at calculations of spontaneous emission rate from this state. For that it must be
\[
\gamma_1'' \sim\Omega_p'\Omega_p''/\Gamma \sim \gamma_r/(k_{LPR}r)^5 \ll \omega_{LPR}.
\]
The non-radiative dumping rate $\gamma_{1nr} = \gamma_1 - \gamma_{1r}$ of state \rf{06} is
\beq
    \gamma_{1nr} = \gamma_{nr} + (\Omega_{p}/\Gamma)^2\Gamma_{nr}. \lb{09}
\eeq
Thus the nanoparticle accelerates the radiation from the emitter and the emitter non-radiative dumping.  This acceleration is described by term $(\Omega_p/\Gamma)^2\Gamma_r$  in Eq. \rf{09}.  Because of $\gamma_r,
 \gamma_{nr} \ll \Gamma$ this term can be large even at weak interaction between the nanoparticle and the emitter, when $(\Omega_{p}/\Gamma)^2 \ll 1$, and the number of excited plasmons is small $n_{pl} \ll 1$.

Dynamics of population $W_1$ of state $\left|\Psi_{11}\right>$ is described by equation
\beq
    \dot{W}_1 = - 2\gamma_1W_1 = - 2\gamma_{1r}W_1-2\gamma_{1nr}W_1. \lb{010}
\eeq
The first term in the right side of Eq.\rf{010} is the power of radiative, the second -- non-radiative losses. Suppose, at $t=0$ the system is excited.  Then a radiation power $P_{11}   = 2\gamma_{1r}W_1= 2\gamma_{1r}\exp{(-2\gamma_1t)}$, the number of spontaneously emitted photons is $\gamma_{1r}/\gamma_{1}$. Here and below the meaning of indexes $N$ and $n$ in $P_{Nn}$ is the same as  indexes in $\left|\Psi_{Nn}\right>$.

Let us take two systems, each of them is  the emitter near the nanoparticle. Suppose the systems are far away from each other and  do not interact. Rate equations describing decay of populations of common states of systems are:
\beqr
    \dot{W}_{11} & = & - 4\gamma_1W_{11} \nonumber\\
    \dot{W}_{10} & = & 2\gamma_1(W_{11} - W_{10}) \lb{011}\\
    \dot{W}_{01} & = & 2\gamma_1(W_{11} - W_{01}), \nonumber
\eeqr
where, for example, ${W}_{10}$ is the population of the state with the first system excited, while the second one is in the ground state etc. Obviously that $W_{10} = W_{01} = \tilde{W}_1$. Suppose both systems are excited at $t=0$, so that the power $P_{11}^{(2)}$ of spontaneous emission from two systems is
\[
    P_{11}^{(2)} = 4\gamma_{1r}W_{11} + 2\gamma_{1r}(W_{10} + W_{01}) = 4\gamma_{1r}(W_{11} + \tilde{W}_1) = 4\gamma_{1r}\tilde{W},
\]
where $\tilde{W} = W_{11} + \tilde{W}_1$. From Eqs.\rf{011} we get  $\dot{\tilde{W}}  =  -2\gamma_1\tilde{W}$, $\tilde{W}(t) = e^{-2\gamma_1t}$, so that
\beq
P_{11}^{(2)} = 2\cdot 2\gamma_{1r}e^{-2\gamma_1t} = 2P_{11}, \lb{012}
\eeq
and the number of emitted photons is $2\cdot(\gamma_{1r}/\gamma_1)$. Thus, as it is expected, two non-interacting systems "emitter+nanoparticle" emit twice more photons with twice higher emission power  and with the same decay rate  as single "emitter+nanoparticle" system. The ratio of radiation powers  $P_{11}^{(2)}(t)/P_{11}(t) = 2$ remains constant.   The efficiency of emission, that is a ratio of the number of emitted photons to the number of initial excitations, is $\gamma_{1r}/\gamma_1$, -- the same for one and for two systems. In the next Section we compare radiative properties for two highly separated and non-interacting "emitter+nanoparticle" systems with properties of single nanoparticle with two emitters near it.

\section{Two emitters near nanoparticle}

Now we consider a system consisted of the metal nanoparticle and two emitters near it, LPR frequency of the nanoparticle is equal to the  emitter transition frequency. Initially, at $t=0$, the system is excited in its first excited state by a resonant pulse of electromagnetic field of certain polarization. After the excitation the system decays to its ground state by spontaneous emission and by non-radiative processes, when some emitted photons are  absorbed in the nanoparticle. Emitters interact with the nanoparticle and with each other through electromagnetic field, coupling constant of interaction of $i=1,2$ emitter with the nanoparticle is $\Omega_{pi}$, we take it real, coupling constant of interaction between emitters is $\Omega_{12} = \Omega_{12}' + i\Omega_{12}''$.  In order to simplify analysis we suppose the same directions of polarizations of initial pulse, transition dipole momentums of emitters and the nanoparticle. It can be, for example,  when emitters and the center of the nanoparticle are on the same line, while polarization of the initial pulse is perpendicular to this line, see Fig.1.
%
%
\begin{figure}[h]
\bc \includegraphics[width=7cm]{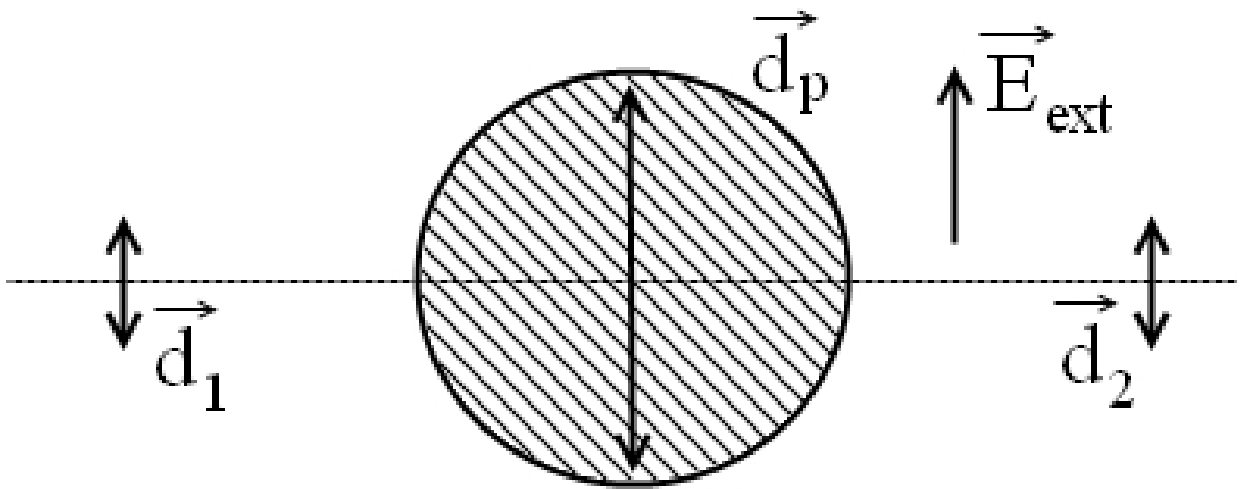}\\
\vspace{0.5cm}\baselineskip 0.5cm \parbox{16cm}{\small  {\bf Fig.1} {Spatial configuration of location and polarization of emitters, nanoparticle and initial pulse with amplitude $\vec{E}_{ext}$. $\vec{d}_{i}$ are dipole momentums of $i=1,2$ emitters, $\vec{d}_p$ is dipole momentum of the nanoparticle.}} \ec
\end{figure}
%
%
We denote common  states of emitters as $\left|\alpha\beta\right>$; $\alpha,\beta =0$ means  ground, $\alpha,\beta =1$ -- excited states; ground and excited states of the nanoparticle are, respectively, $\left|0\right>$ and $\left|1\right>$.   Wavefunction $\left|\Psi_{Nn}\right>$ of the state with $n=2$ excitations of the system with $N=2$ emitters and the nanoparticle is
\beq
    \left|\Psi_{22}\right> = C_{110}\left|11\right>\left|0\right> + C_{101}\left|10\right>\left|1\right> +
    C_{011}\left|01\right>\left|1\right>,      \lb{1}
\eeq
wavefunction of the state with one excitation is
\beq
    \left|\Psi_{21}\right> = C_{100}\left|10\right>\left|0\right> + C_{010}\left|10\right>\left|0\right> +
    C_{001}\left|00\right>\left|1\right>.      \lb{2}
\eeq
Here $C_{\alpha\beta\gamma}$ is  probability amplitude, indexes $\alpha,\beta,\gamma = 0,1$,  the last index $\gamma$ corresponds to the state of the nanoparticle. Only states $\left|\alpha\beta\right>\left|\gamma\right>$ with the same number of excitations interact with each other in resonance, close sets of equations can be written for their probability amplitudes. For $n=2$ excitations:
\beqr
    \dot{C}_{110} & = & -2\gamma C_{110} -i\Omega_{p1}C_{011}-i\Omega_{p2}C_{101} \lb{3}\\
    \dot{C}_{101} & = & -(\gamma + \Gamma)C_{101} -i\Omega_{12}C_{011}-i\Omega_{p2}C_{110} \lb{4}\\
    \dot{C}_{011} & = & -(\gamma + \Gamma)C_{011} -i\Omega_{12}C_{101}-i\Omega_{p1}C_{110}, \lb{5}
\eeqr
for $n=1$ excitation:
\beqr
    \dot{C}_{100} & = & -\gamma C_{100} -i\Omega_{p1}C_{001}-i\Omega_{12}C_{010} \lb{6}\\
    \dot{C}_{010} & = & -\gamma C_{010} -i\Omega_{12}C_{100}-i\Omega_{p2}C_{001} \lb{7}\\
    \dot{C}_{001} & = & -\Gamma C_{001} -i\Omega_{p1}C_{100}-i\Omega_{p2}C_{010}, \lb{8}
\eeqr
the meaning of $\gamma$ and $\Gamma$ is the same as in Section 1.

Let us suppose that the distance between an emitter and the nanoparticle is the same, therefore $\Omega_{p1} = \Omega_{p2} = \Omega_{p}$. Then we introduce $C_{2\pm} = C_{011}\pm C_{101}$, $C_{1\pm} = C_{010}\pm C_{100}$ and write instead of Eqs. \rf{3} -- \rf{5} and Eqs. \rf{6} -- \rf{8}:
\beqr
    \dot{C}_{110} & = & -2\gamma C_{110} -i\Omega_{p}C_{2+} \lb{15}\\
    \dot{C}_{2+} & = & -(\gamma + \Gamma + i\Omega_{12})C_{2+} -2i\Omega_{p}C_{110} \lb{16}\\
    \dot{C}_{2-} & = & -(\gamma + \Gamma - i\Omega_{12})C_{2-} \lb{17}
\eeqr
and
\beqr
    \dot{C}_{1+} & = & -(\gamma +  i\Omega_{12})C_{1+} -2i\Omega_{p}C_{001} \lb{18}\\
    \dot{C}_{1-} & = & -(\gamma -  i\Omega_{12})C_{1-}  \lb{19}\\
    \dot{C}_{001} & = & -\Gamma C_{001} -i\Omega_{p}C_{1+}. \lb{20}
\eeqr
It is shown in \ct{4}, that in the short-distance limit $\Real{(i\Omega_{12})} = \gamma_r$, so that
\beqr
    \dot{C}_{110} & = & -2\gamma C_{110} -i\Omega_{p}C_{2+} \lb{21}\\
    \dot{C}_{2+} & = & -(2\gamma_r + \gamma_{nr} + \Gamma + i\Omega'_{12})C_{2+} -2i\Omega_{p}C_{110} \lb{22}\\
    \dot{C}_{2-} & = & -(\gamma_{nr} + \Gamma - i\Omega'_{12})C_{2-} \lb{23}
\eeqr
and
\beqr
    \dot{C}_{1+} & = & -(2\gamma_r + \gamma_{nr} +  i\Omega'_{12})C_{1+} -2i\Omega_{p}C_{001} \lb{24}\\
    \dot{C}_{1-} & = & -(\gamma_{nr} -  i\Omega'_{12})C_{1-}  \lb{25}\\
    \dot{C}_{001} & = & -\Gamma C_{001} -i\Omega_{p}C_{1+}. \lb{26}
\eeqr
We suppose fast relaxation of plasmon, i.e. condition \rf{003}, and adiabatically eliminate ${C}_{2\pm}$ and $C_{001}$ from Eqs. \rf{22} \rf{23} and \rf{26} by setting there time derivatives to zero; after that we write instead of Eqs.\rf{21} -- \rf{23} and \rf{24} -- \rf{26}:
\beqr
    \dot{C}_{110} & = & -2(\gamma + \Omega_{p}^2/\Gamma)C_{110}  \lb{27}\\
    C_{2+} & = &  -(2i\Omega_{p}/\Gamma)C_{110} \lb{28}\\
    {C}_{2-} & = & 0, \lb{29}
\eeqr
and
\beqr
    \dot{C}_{1+} & = & -(2\gamma_r + \gamma_{nr} + 2\Omega_p^2/\Gamma + i\Omega'_{12})C_{1+}  \lb{30}\\
    \dot{C}_{1-} & = & -(\gamma_{nr} -  i\Omega'_{12})C_{1-}  \lb{31}\\
    {C}_{001} & = & -i(\Omega_{p}/\Gamma)C_{1+}. \lb{32}
\eeqr
Terms $\sim i\Omega'_{12}$ in Eqs.\rf{30}, \rf{31} describes changes of state's energies due to the interaction between emitters.

Similar to the case of single emitter near the nanoparticle we find wavefunction $\left|\Psi_{22}\right> = \left|\psi_{22}\right>e^{-i\gamma_{22}t}$ from Eqs.\rf{30} -- \rf{32} with
\beq
    \left|\psi_{22}\right> = \left|11\right>\left|0\right> - \frac{i\Omega_p}{\Gamma}\left(\left|10\right> +
    \left|01\right>\right)\left|1\right>,      \lb{34}
\eeq
\beq
    \gamma_{22} = 2(\gamma + \Omega_{p}^2/\Gamma).    \lb{33}
\eeq
The state \rf{34} contains small, with the height $\sim (\Omega_{p}/{\Gamma})^2 \sim \gamma_r/[\Gamma(k_{LPR}r)^{6}] \ll 1$, contribution 
of $(1/\sqrt{2})\left(\left|10\right> + \left|01\right>\right)\left|1\right>$ symmetric state of emitters and excited nanoparticle. In spite of the probability to find this state is small, the dipole momentum of transition from such state is large, so that this state may give high contribution to the emission.  The increment $\gamma_{22}$ describes full decay (radiative and non-radiative) of the state to lower energy state.
Solving Eqs.\rf{30} -- \rf{32} we find two states:  $\left|\Psi_{21}^{(\pm)}\right> = \left|\psi_{21}^{(\pm)}\right>e^{-i\gamma_{21}^{(\pm)}t}$,
\beqr
    \left|\psi_{21}^{(+)}\right> & = & \frac{1}{\sqrt{2}}\left[\left(\left|10\right> + \left|01\right>\right)\left|0\right>
    - \frac{2i\Omega_{p}}{\Gamma}\left|00\right>\left|1\right>\right]    \lb{36}\\
    \left|\psi_{21}^{(-)}\right> & = & \frac{1}{\sqrt{2}}\left(\left|10\right> - \left|01\right>\right)\left|0\right>,   \lb{37}
\eeqr\beq
    \gamma_{21}^{(+)} = 2\gamma_r + \gamma_{nr} + 2\Omega_p^2/\Gamma, \hspace{1cm} \gamma_{21}^{(-)} = \gamma_{nr}.  \lb{38}
\eeq
Thus the relaxation of two excited states of emitters near the nanoparticle occurs through symmetric and antisymmetric states of emitters:
\beq
    \Psi_{22} \rightarrow  \Psi_{21}^{(+)} \rightarrow \Psi_{20}, \hspace{1cm}
    \Psi_{22} \rightarrow  \Psi_{21}^{(-)} \rightarrow \Psi_{20},   \lb{39}
\eeq
where $\Psi_{20}$ is a background state of the system. Population decay rates for each of two first transitions in \rf{39} are  $\gamma_{22}^{(+)} \equiv 2(\gamma_{22} - \gamma_{nr}) = 2\gamma_{21}^{(+)} \equiv 2\gamma_+$,  population decay rate of each of two second transitions in \rf{39} is
$2\gamma_{nr} \equiv 2\gamma_-$. 

Note that states \rf{36}, \rf{37} contains fully symmetric and anti-symmetric states of emitters only at the assumption of equal interaction of each emitter with the nanoparticle. Fluctuations in the emitter-nanoparticle interaction can reduce or even destroy the symmetry and therefore superradiance, the role of such fluctuations has to be investigated specially. Here we did not neglected by the interaction of emitters with each other, such interaction leads to two times increase the radiation rate: $\gamma_r \rightarrow 2\gamma_r$ of symmetric state of emitters and cancelation of the radiation from  anti-symmetric state of emitters $\gamma_r \rightarrow 0$  in Eqs.\rf{22} and \rf{25} correspondingly. This the interaction between emitters provides full accordance of radiation rates in Eqs.\rf{22} and \rf{25} with results of calculations of radiation rates basing at full dipole matrix elements presented below.

In order to describe the radiation one has to extract radiation relaxation rates from full rates $\gamma_+$.  For that we find matrix elements of full dipole momentum operator $\hat{D}_2 = \hat{d}_1+\hat{d}_2+\hat{d}_p$ for the first group of transitions in \rf{39}, here $\hat{d}_i$, $i=1,2$ is the operator of dipole momentum of transition of $i$-th emitter with real matrix element $d$. We calculate
\[
\left<\psi_{22}\right|\hat{D}_2\left|\psi_{21}^{(+)}\right> = \]\[
    \left\{\left<0\right|\left<11\right| - \frac{i\Omega_p}{\Gamma}\left<1\right|\left(\left<10\right| +
\left<01\right|\right)\right\}(\hat{d}_1+\hat{d}_2+\hat{d}_p)\frac{1}{\sqrt{2}}\left\{\left(\left|10\right> +
    \left|01\right>\right)\left|0\right>
    - \frac{2i\Omega_{p}}{\Gamma}\left|00\right>\left|1\right>\right\} = \]\beq
    \sqrt{2}\left(d - \frac{i\Omega_{p}}{\Gamma}d_p\right).  \lb{40}
\eeq
The radiative rate for this transition is $\sim \left|\left<\hat{D}_2\right>\right|^2$. Therefore radiative part $\gamma_{+r}$ of $\gamma_+$ is
\beq
    \gamma_{+r} = 2\left[ \gamma_r + (\Omega_{p}/{\Gamma})^2\Gamma_r \right].      \lb{41}
\eeq
The non-radiative rate of $\Psi_{22} \rightarrow  \Psi_{21}^{(+)}$  transition is $\gamma_+  - \gamma_{+r}$.
Now we calculate the radiation rate for $\Psi_{21}^{(+)} \rightarrow \Psi_{20}$ transition. The matrix element is
\[
\left<\psi_{00}\right|\hat{D}_2\left|\psi_{21}^{(+)}\right> = \]\beq
\left<0\right|\left<00\right|(\hat{d}_1+\hat{d}_2+\hat{d}_p)\frac{1}{\sqrt{2}}\left[\left(\left|10\right> +
    \left|01\right>\right)\left|0\right>
    - \frac{2i\Omega_{p}}{\Gamma}\left|00\right>\left|1\right>\right] =
    \sqrt{2}\left(d - \frac{i\Omega_{p}}{\Gamma}d_p\right),  \lb{43}
\eeq
the same as the one given by Eq.\rf{40}. 
The scheme of relaxation of populations of states of two emitters and the nanoparticle is shown in in Fig.2.
%
%
\begin{figure}[h]
\bc \includegraphics[width=7cm]{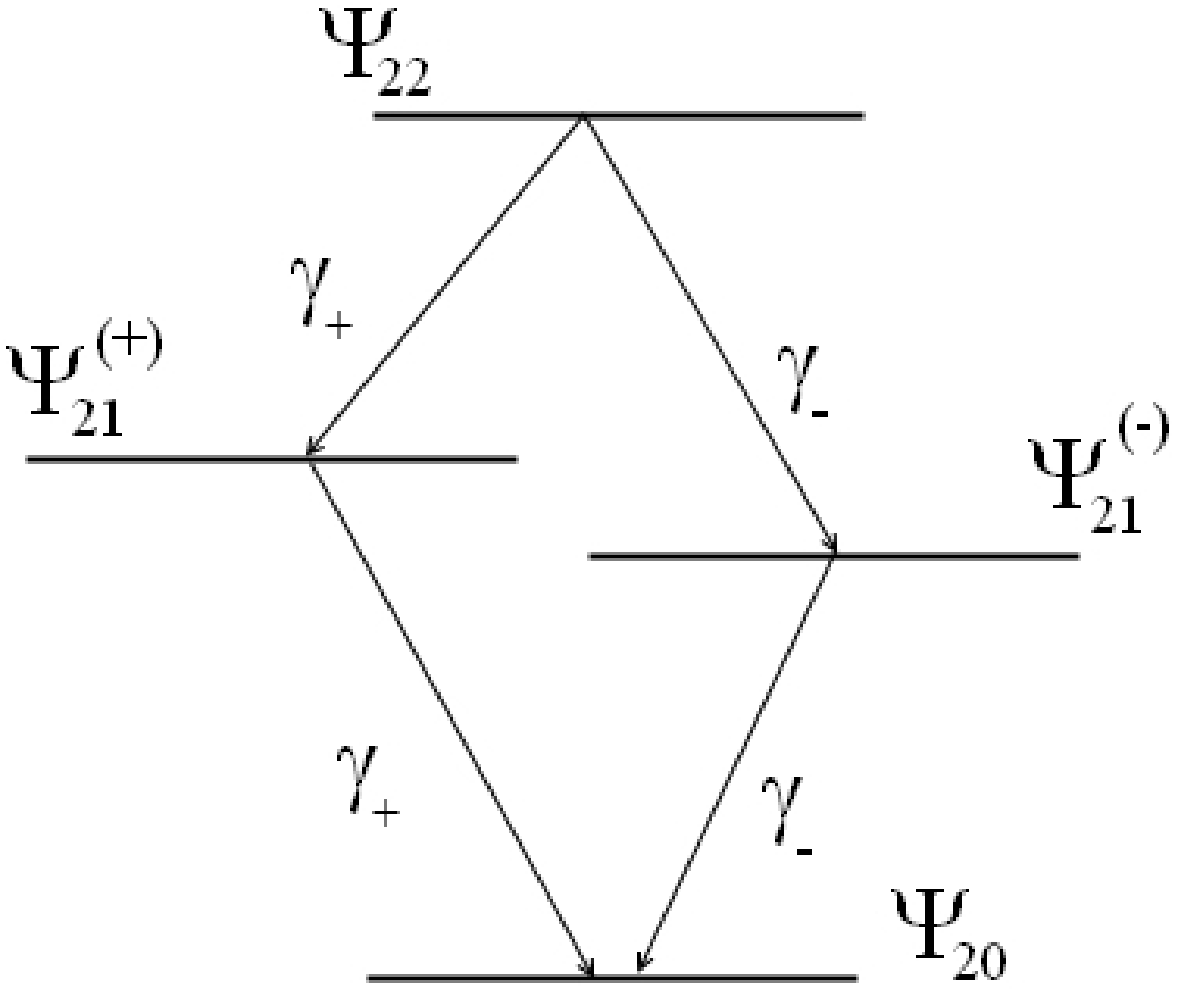}\\
\vspace{0.5cm}\baselineskip 0.5cm \parbox{16cm}{\small  {\bf Fig.2} {States and relaxation of two emitters near the nanoparticle.}} \ec
\end{figure}
%
%
One can write the set of rate equations for populations $W_{22}$, $W_{21}^{(+)}$ of states $\Psi_{22}$ and $\Psi_{21}^{(+)}$ shown in Fig.2.
\beqr
    \dot{W}_{22} & = & -2(\gamma_+ + \gamma_-)W_{22}  \lb{46}\\
    \dot{W}_{21}^{(+)} & = & 2\gamma_+(W_{22} - {W}_{21}^{(+)}).  \lb{47}
\eeqr
The radiation power of the system
\beq
            P_{22} = \gamma_{+r}(W_{22} + {W}_{21}^{(+)}), \lb{48}
\eeq
will be estimated in the next Section.

\section{Estimations of radiation power}

We take the same values of parameters as in \ct{5,6}:  gold spherical nanoparticle of radius $a=19$~nm, this radius corresponds to the maximum LPR quality factor \ct{6}. Suppose, that the nanoparticle has a dielectric shell, for example, silica shell, emitters are on the surface of the shell. Silica index of refraction is $n_{Si} = 1.5$, the nanoparticle is in water with refractive index $n_0 = 1.33 \approx n_{Si}$. LPR wavelength in vacuum is $\lambda_{LPR} = 525$~nm, LPR frequency $\omega_{LPR} = 3.59\cdot 10^3$~THz, LPR Q-factor $Q=34$ \ct{6}, so that full LPR half-width $\Gamma = (\pi c_0)/(\lambda Q) \approx 53$~THz;   $\Gamma_{nr} \approx \Gamma_r$ corresponds to maximum $Q$: $\Gamma_{nr} = \Gamma_r = \Gamma/2 = 26.5$~THz. Well-known expressions for $\gamma_r$, $\Gamma_r$ and expression for $\Omega_p$ followed from Eqs.(21) and (22) of \ct{7} in the short-distance limit for configuration as in Fig.1, are:
\[
    \Omega_p = -\frac{n_0\omega_{LPR}^3d_ed_p}{\hbar c_0^3(k_{LPR}r)^3}, \hspace{0.5cm}
    \gamma_r = \frac{2n_0d_e^2\omega_{LPR}^3}{3\hbar c_0^3},\hspace{0.5cm}\Gamma_r = \frac{2n_0d_p^2\omega_{LPR}^3}{3\hbar c_0^3},
\]
so that
\[
    \Omega_p^2 = \frac{9\gamma_r\Gamma_r}{4(k_{LPR}r)^6}.
\]
According with  \ct{07}
\[
    \gamma_{nr} = \frac{\gamma_r}{2(r/a -1)^3}.
\]
Therefore we can write for relaxation rates:
\[
    \frac{\gamma_+}{\gamma_r} = 2 + \frac{1}{2(r/a -1)^3} + \frac{9}{4(k_{LPR}a)^6(r/a)^6},\hspace{1cm}\frac{\gamma_-}{\gamma_r} =\frac{1}{2(r/a -1)^3}.
\]
If we normalize time to $(2\gamma_r)^{-1}$, then for given $k_{LPR}a = 2\pi n_0a/\lambda_{LPR} = 0.34$ we have single free parameter $r/a$ in equations \rf{46}, \rf{47}  and in expression \rf{48}. Solving \rf{46}, \rf{47} we came to
\beqr
    {W}_{22} & = & e^{-2({\gamma}_+ + {\gamma}_-)t}  \lb{53}\\
    {W}_{21}^{(+)} & = & ({{\gamma}_+}/{{\gamma}_-})
    \left(1-e^{-2{\gamma}_-t}\right)e^{-2{\gamma}_+t}.  \lb{54}
\eeqr
Radiation power is
\beq
    P_{22}(t) = 2{\gamma}_{+r}\left[e^{-2{\gamma}_-t} + \frac{{\gamma}_+}{{\gamma}_-}
    \left(1-e^{-2{\gamma}_-t}\right)\right]e^{-2{\gamma}_+t}, \lb{55}
\eeq
total number of emitted photons --
\beq
            \int_0^{\infty}P_{22}(t)dt = 2\frac{{\gamma}_{+r}}{{\gamma}_+ + {\gamma}_-} = 2\frac{{\gamma}_{1r}}{{\gamma}_1},   \lb{56}
\eeq
that is the number 2 of initial excitations in the system multiplied by the radiation efficiency ${{\gamma}_{1r}}/{{\gamma}_1}$, which is the same as for single "emitter+nanoparticle" system or for two non-interacting "emitter+nanoparticle" systems.

Figure 3 shows, for various $r/a$, radiation power $P(t)$ in photons per second for system of two emitters and the nanoparticle  -- solid curves 1, 2 and 3, and for two emitters without the nanoparticle -- dashed curves: when emitters are close to each other -- curve 4, for large distance between emitters -- curve 5. Curve 4 corresponds to the power of radiation given by Eq.\rf{55} in the limit $\gamma_- \rightarrow 0$, $\gamma_+, \gamma_{+r} \rightarrow \gamma_r$, radiation power for curve 5 is $P(t) = 4\gamma_re^{-2\gamma_r t}$. From Fig.3 we see, that the nanoparticle noticerably accelerates collective spontaneous emission,  even when the distance between the particle and the emitter is relatively large $r/a > 2$. Efficiencies of the radiation are 0.6, 0.73 and 0.91 for $r/a = 2.5$, $3$ and $4$, respectively.
%
%
\begin{figure}[h]
\bc \includegraphics[width=12cm]{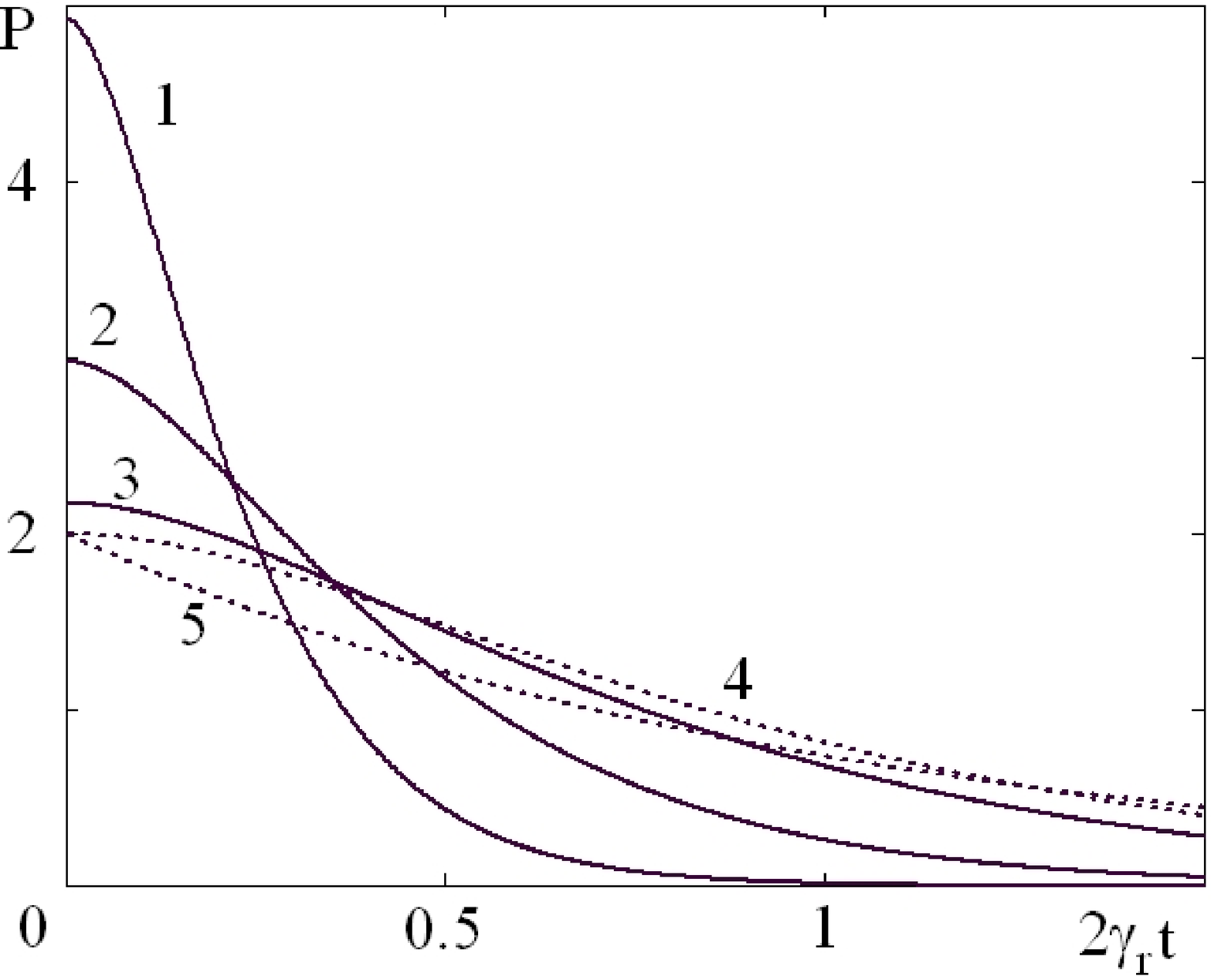}\\
\vspace{0.5cm}\baselineskip 0.5cm \parbox{16cm}{\small  {\bf Fig.3} {Radiation power in photons per second for two emitters near gold nanoparticle of radius $a=19$~nm for various distances $r$ between an emitter and the center of nanoparticle: $r/a = 2.5$, $3$ and $4$ -- curves 1 -- 3. Curve 4 is for radiation of two close emitters to each other without the nanoparticle, curve 5 -- for two such emitters far away from each other.}} \ec
\end{figure}
%
%

One can see that collective spontaneous emission from two emitters near the nanoparticle is faster that from two highly separated and non-interacted systems "nanoparticle+emitter" from Fig.4.
%
%
\begin{figure}[h]
\bc \includegraphics[width=12cm]{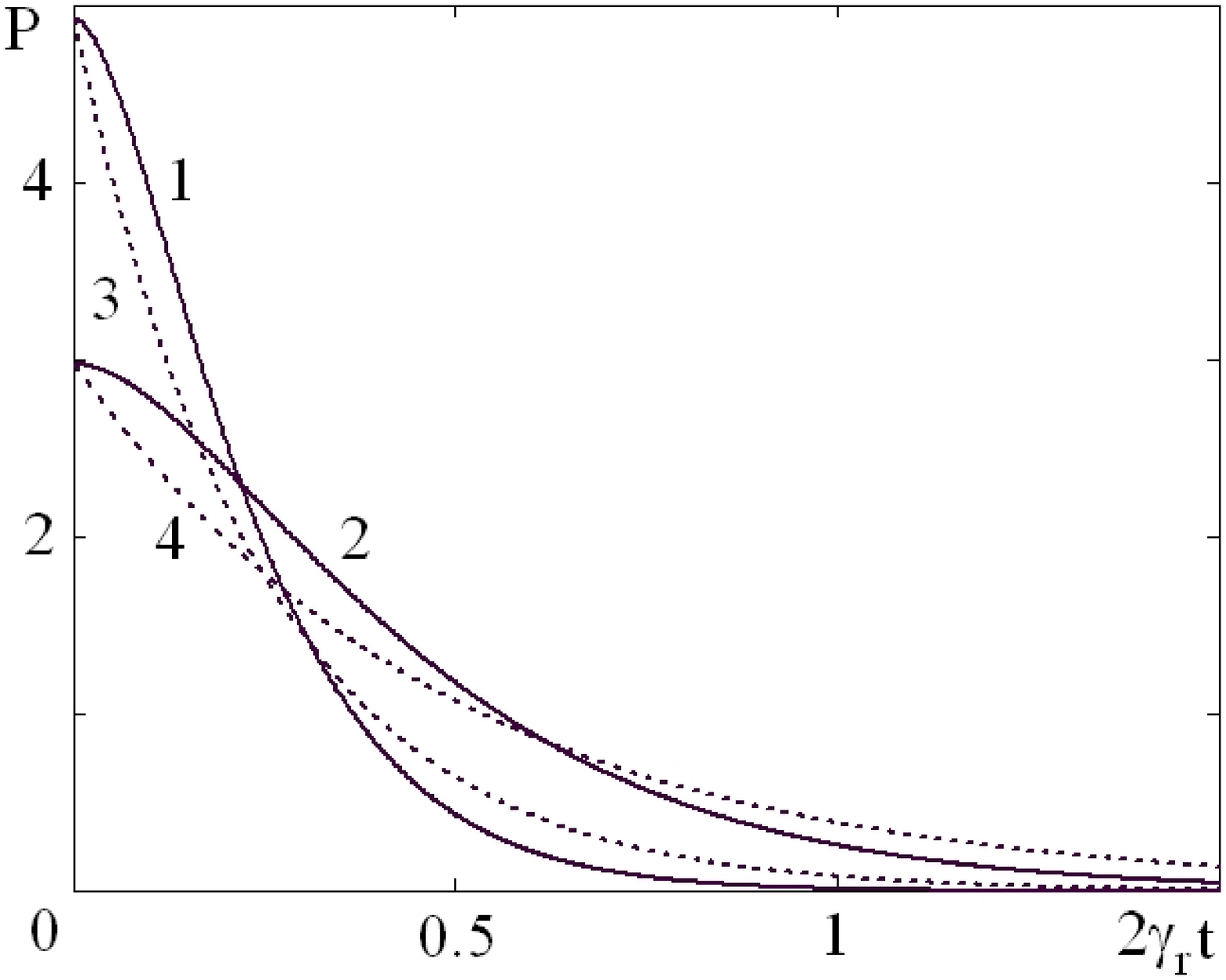}\\
\vspace{0.5cm}\baselineskip 0.5cm \parbox{16cm}{\small  {\bf Fig.4} {Radiation power of two emitters near the nanoparticle for $r/a = 2.5$ and $3$: solid curves 1 and 2, respectively; and for two highly separated "nanoparticle+emitter" systems -- dashed curves 3 and 4 for $r/a = 2.5$ and $3$, respectively.}} \ec
\end{figure}
%
%
We note that there is no maximum in the collective spontaneous emission of two emitters near the nanoparticle at $t>0$, -- similar to the case of two emitters without a nanoparticle \ct{4}. However at some time interval after $t=0$ the power of collective spontaneous emission is greater than the power of spontaneous emission from separated and non-interacting systems. Fig.5 shows ratios of collective spontaneous emission powers to powers of emission from two separated systems for $r/a = 2$, $2.5$ and $3$ -- solid curves, and the ratio of emission powers from two close and two separated emitters without a nanoparticle -- curve 4. It is interesting to note that all maxima of such ratios have the same values. The nanoparticle accelerates the radiation. Maxima of power ratios appear as shortly after $t=0$ as smaller is $r/a$. After some time  power ratios became smaller than 1. As closer emitters are to the nanoparticle as smaller is the time  necessary for emission of all photons.
%
%
\begin{figure}[h]
\bc \includegraphics[width=12cm]{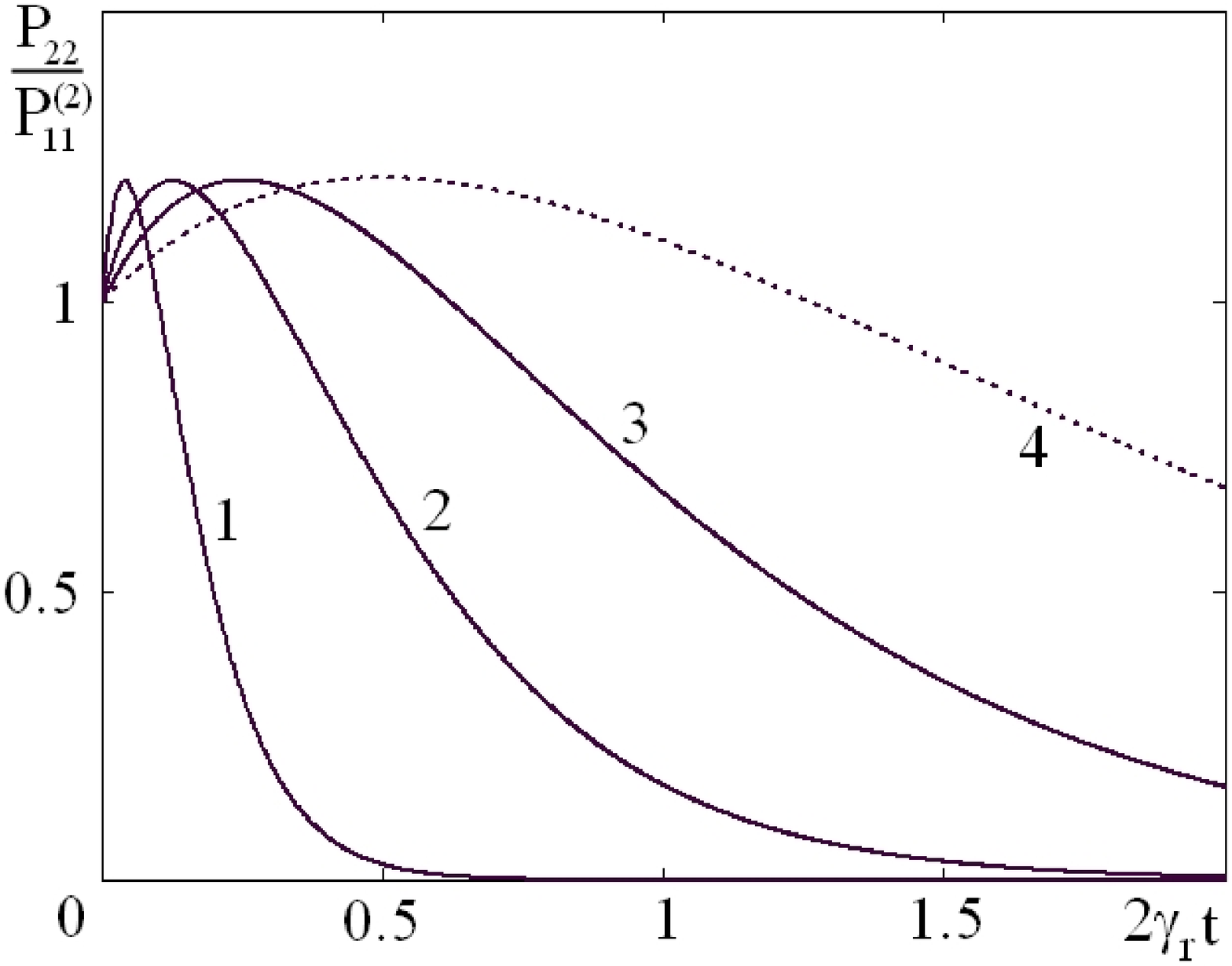}\\
\vspace{0.5cm}\baselineskip 0.5cm \parbox{16cm}{\small  {\bf Fig.5} {Ratio of emission power $P_{22}$ of two emitters near the nanoparticle (Eq.\rf{55}) to the power of emission $P_{11}^{(2)}$ of two independent "nanoparticle+emitter" systems (Eq.\rf{012}) for $r/a = 2$, $2.3$ and $3$ -- solid curves 1,2 and 3, respectively. Dashed curve 4 is for two emitters without the nanoparticle.}} \ec
\end{figure}
%
%
\section{Conclusion}
We presented a method of quantum-mechanical analysis of radiation of resonant emitters near metal nanoparticle. Our approach is based on Schrodinger picture, i.e. we use wave functions. We show how to obtain wave functions in the case, when a part of the system has strong dissipation, here such part is a metal nanoparticle. Our approach will be useful for analysis of various dissipative quantum systems.   

We have shown that resonant radiation from two emitters, interacting with each other near metal nanoparticle, is occurred through symmetric states of emitters. Non-symmetric states do not radiate. Such radiating and non-radiating states are similar with states in Dicke model of superradiance without the nanoparticle. In a difference with Dicke model  non-radiative dissipation presents in our system and it is taken into account. Due to non-radiative processes non-symmetric (dark) states of the system are populated, some part of population of symmetric (bright) states also decays due to non-radiative processes. 

Non-radiative decay, obviously, reduces the radiation. If emitters are not too close to the nanoparticle surface -- at the distance of the order of the nanoparticle radius, then about a half of the energy stored in emitters is radiated, the rest of the energy is absorbed by the nanoparticle. The nanoparticle accelerates collective spontaneous emission of emitters similar as it accelerates the radiation of single emitter. Assumption that the interaction of any emitter with  nanoparticle is the same -- important condition necessady for formation of purely symmetric and non-symmetric states of emitters. Different interaction of the nanoparticle with different emitters brakes the symmetry of states. The question of how strong such difference influences collective spontaneous emission is a subject of special research.

Approach presented above can be generalized straightforwardly to the case of many emitters near the nanoparticle. Then, supposing approximately equal interaction of emitters and the nanoparticle, one can assume that the radiation will be through symmetric states of emitters, as in \ct{06}, but even without strong interaction between emitters.   The non-radiative decay of emitters through non-symmetric states, neglected in \ct{06}, gives relatively small contribution, if emitters are not too close to the nanoparticle surface and the nanoparticle sufficiently accelerates the radiation of emitters. The efficiency of the superradiance will be about the efficiency of radiation of single emitter near the nanoparticle, that is $\sim 50\%$.

Here particular spatial configuration of the particle and emitters  was considered: when polarization of the nanoparticle, transitions of emitters and of the pulse of excitation coinside: this is, in fact, a one-dimentional case with coherent excitation. More general, 3-dim case, when dipole momenta of any direction may appear at the radiation, can be considered similar way, however the analysis of such cases are cumbersome and we leave it for the future. One can also take into consideration continues incoherent pump of emitters and to analyse a supperradiance lasing with plasmonic nanoparticle. The paper \ct{08} points out to growing interest to such superradiance lasers.

There is an important question: what is the number of emitters near the nanoparticle necessary for the maximum efficiency of superradiance? The well-known viewpoint is that inhomogeneous broadening due to the interaction of emitters with each other destroys superradiance in small volumes $\ll \lambda_{LPR}^3$ \ct{3a}. Thus if the number of emitters in the nanoparticle shell is too large, the interaction between them will be so strong that the inhomogeneous broadening destroys the superradiance. We note, however, that the inhomogeneous broadening is caused by {\em fluctiations} in the interaction of an emitter with others rather than by the interaction itself. Indeed, in perfectly ordered structure of emitters no fluctuations exist and therefore no inhomogeneous broadening of emitter transitions presents, though the emitter transition frequency is different from the one of isolated emitter. We can't organize perfectly ordered ensemble of emitters in the nanoparticle shell, however for sufficient number of emitters fluctuations in their interactions can be much smaller than the mean value of the interaction between them. Besides, following \ct{3a} we suggest,  that the inhomogeneous broadening sufficiently influences superradiance if the inhomogeneous linewidth is larger then $1/t_s$, where $t_s$ is the time of superrdiance. However $t_s$ near the nanoparticle is much shorter than without it \ct{06}. Thus larger inhomogeneous broadenind is necessary for distruction of the superradiance with the nanoparticle than without it. Though the above arguments are phenomenological ones, they give us a hope to find conditions for efficient superradiance near plasmonic nanoparticle and, may be, even to develope plasmonic superradiance nano-emitter or nano-laser. 

\end{document}